\documentclass[doublecol]{epl2} 

\newcommand{\ud}{{\rm d}}
\newcommand{\RH}{\mathrm{RH}}

\newcommand{\edot}{{\dot{E}}}

\newcommand{\VT}{V_{\rm T}}
\newcommand{\mdry}{m_{\rm dry}}
\newcommand{\rhos}{\rho_{\rm silica}}

\usepackage{graphicx,chemist}
\usepackage{subfigure}
\usepackage{epsfig}
\usepackage{amssymb,amsmath}
\usepackage{epstopdf}
\usepackage{soul}

\graphicspath{{Figures/}}
\begin{document}
 
\title{Role of evaporation rate on the particle organization and crack patterns obtained by drying  a colloidal layer}
\shorttitle{Evaporation rate effects during colloidal drying}

\author{K. Piroird\inst{1,2} \and V. Lazarus\inst{1}\thanks{E-mail: \email{veronique.lazarus@u-psud.fr}} \and G. Gauthier\inst{1} \and A. Lesaine\inst{1,2} \and D. Bonamy\inst{2} \and C.~L. Rountree\inst{2}}
\shortauthor{K. Piroird \etal}

\institute{  
\inst{1} Laboratoire FAST, Univ. Paris-Sud, CNRS, Universit\'e Paris-Saclay, F-91405, Orsay, France.\\
\inst{2}  SPEC, CEA, CNRS, Universit\'e Paris-Saclay, 91191 Gif-sur-Yvette, France. \\
}

\begin{abstract}
{
A scientific hurdle in  manufacturing solid films by drying colloidal layers is preventing them from fracturing. This paper examines how the drying rate of colloidal liquids influences the particle packing at the nanoscale in correlation with the crack patterns observed at the macroscale. Increasing the drying rate results in more ordered, denser solid structures, and the dried samples have more cracks.Yet, introducing a holding period (at a prescribed point) during the drying protocol results in a {more} disordered solid structure with significantly less cracks. To interpret these observations, this paper conjectures  that a longer drying protocol favors the formation of aggregates. It is further argued that the number and size of the aggregates increase as the drying rate decreases.  This results in the formation of a more disordered, porous film from the viewpoint of the particle packing, and a more resistant film, i.e. less cracks, from the macroscale viewpoint.}
\end{abstract}


\maketitle

Obtaining solid layers via drying of colloidal suspensions is central to many technological fields \cite{MujXuYu09}:Printing and painting \cite{Rou13}, manufacturing  protective or decorative coatings \cite{KedRou10}, designing materials at the nanoscale by low-cost processes \cite{MarGelLoh11, LenMerSal12}\ldots. From a fundamental {viewpoint}, colloidal systems provide model systems to mimic the atomic behavior at a larger, more accessible length-scale \cite{YetBla03, NatMat14}.Understanding, predicting and controlling the mechanisms driving their self-assembly  represents a major challenge. Obtaining this control will aid in ensuring the resistance to failure while staying eco-friendly \cite{KedRou10}, especially for applications requiring thick and/or hard coatings. Amongst several other parameters (thickness \cite{LazPau11desc}, ionic strength \cite{AllPauPar99}, substrate adhesion \cite{GroKap94}...), the drying rate significantly impacts the cracking patterns  \cite{CadHul02, GauLazPau10, BouGioPau14}. This observation remains poorly understood for several reasons. During drying, mass loss  plays simultaneously\cite{GauLazPau10} on the evolution of both the loading applying externally to the layer \cite{Sch89, CheLaz13} and the intrinsic material properties of the forming solid \cite{GioPau14}. From a macroscopic point of view, these two effects can hardly be deconvoluted. An important question  to clarify is whether the drying rate plays only on the formation kinetics  yielding the same material or  if it  changes the way the material self-structures during evaporation leading to variation in the  final state.

This Letter experimentally investigates how the evaporation kinetics influences the material properties at the particle (nano)scale and the fracture patterns {at the continuum scale}. For this purpose, a colloidal suspension of hard monodisperse nanospheres was dried in a controlled atmosphere of tunable humidity. An atomic force microscope (AFM) captures the particle arrangement of the final layer topography. Bulk measurements of the packing density complement these surface analyses. Counter-intuitively \cite{MarGelLoh11}, these systems exhibit a progressive transition from disorder to order when increasing the drying rate. This transition goes along with an increase of density and more cracks. The balance between particle convection and diffusion alone cannot explain this transition, and particle agglomeration should be considered. Longer drying protocols {enhance  fracture resistance and} favors particle agglomeration, and consequently disordered packing  structure and porosity.

{\it Setup--} Drying experiments herein invoke model colloidal suspensions of monodisperse silica nanospheres (Ludox HS-40: $\phi^m_{SiO_2}=40\pm 1\%$  in mass of \chemform{SiO_2} colloids with a spec nominal diameter of $12\un{nm}$; $\phi_{Na_2O}^m=0.42\pm 0.04\%$ in mass of free alkalinity as \chemform{Na_2O}; and spec density $\rho_{ludox}=1.30\pm 0.01\un{g/cm}^3$). Experiments herein use the same bottle of suspension\cite{Test} and occur over a restricted time period ($\sim$9 months). This ensures a minimal variation and aging of the initial suspension. A glass Petri dish (inner radius $R=3.5 \un{cm}$) initially contains  $25\un{g}$ ($m_0$) of Ludox. To capture the in-situ mass loss, the Petri dish rests on a precision scale (Sartorius Cubis series, accuracy $10^{-5}\un{g}$). The layer dries uniformly and forms a flat surface, except in the vicinity of the dish edges where an upwards meniscus prevents the drop singularity (hence the coffee-ring type lateral drying \cite{BakDeeDup97}). The layer's initial and final thicknesses are $h_0\simeq 5\un{mm}$ and $h_1\simeq 1.8 \un{mm}$, respectively. An external humidity control system driven by Labview ensures a prescribed constant relative humidity (RH) in the scale housing \cite{BodDouGue10}, which is sealed with a gloverbox quality putty to minimize leaks.

Labview records the suspension's mass  ($m$), the temperature ($T$), and the relative humidity ($\RH$) during the whole drying process. For all the experiments, the temperature was $T=25\pm 2\,^\circ\mathrm{C}$. Hence, $\RH$ controls the  drying rate. Initially, the mass decreases linearly with time (i.e. $\ud m / \ud t\sim constant$) for a prescribed $RH$, as reported in the literature \cite{BriSch90}. Transforming the mass loss into the evaporation rate, $\edot_0$ ($\edot_{0} \equiv (\ud m / \ud t)/(\rho_{water}\pi R^2$)), one finds that $\edot_0$ decreases linearly with the applied $\RH$. Experiments verified that  $\edot_0$ for pure water and Ludox are equal barring the same drying conditions. Modulating $\RH$ from $10\%$ to $95\%$ makes $\edot_{0}$ vary over a decade from $\edot_0=36.1\pm0.7\un{nm/s}$ to $\edot_0=3.4\pm0.2\un{nm/s}$.

\begin{figure}[h!]
\includegraphics[width=\columnwidth]{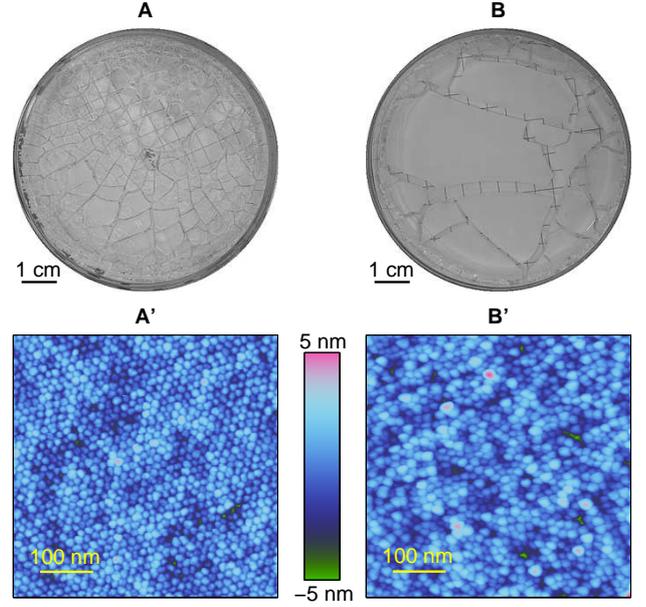}
\caption{Effect of the drying rate at the macro and nanoscales. Top: Final crack patterns observed after evaporation for $\RH=10\%$ (A) and $\RH=95\%$ (B). Bottom: Topographical AFM images depicting the colloid arrangement at the surface of the layers for $\RH=10\%$ (A') and $\RH=95\%$ (B'). The scan size is $500 \times 500\un{nm}^2$.  The out-of-plane height ranges over $10\un{nm}$ which is less than a colloid diameter.
}
\label{fig:rough}
\end{figure}

After the constant evaporation rate regime, a falling-rate regime occurs where the evaporation slows down and eventually stops. Once $\ud m / \ud t$ is negligible
(after $\sim 50$h for $\RH=10\%$ and after $\sim 700$h for $\RH=95\%$), the sample is brought back to ambient humidity. Postmortem observations reported hereafter are independent of the returning protocol:  An abrupt return (obtained by opening the housing) or progressive one (obtained by  bringing the $\RH$ inside the box  back to ambient humidity in a stepwise process as slow as 10\% humidity steps every $12\un{h}$) does not alter the solid porosity nor the particles surface arrangement.
It has also been checked that increasing the Petri dish radius from $2.5$ to $6.8\un{cm}$, keeping the same initial thickness, does not alter experimental observations reporter henceforth.

A camera (USB2 uEye from IDS imaging), located above the scale's enclosure, images the evolution of the sample. As the solvent evaporates, the particles move closer to one another forming a solid network which progressively retracts. The rigid substrate (i.e. the Petri dish) hampers the retraction leading to tensile stresses which cause the layer to fracture (Fig. \ref{fig:rough}). These cracks begin to appear just as the evaporation rate enters the falling-rate regime. This point also coincides with the time at which the  meniscus forms at the top of the particles layer, yielding a decreasing capillary pressure in the pores \cite{CorDufGre03,LazPau11desc,CheLaz13}. A Bruker Dimension Icon Atomic Force Microscope (AFM) images the particle arrangement on the evaporation surface of the fragmented morsels. The AFM records topographical images in Tapping mode using a Bruker RTESPA tip (radius of curvature $\sim 8\un{nm}$). The scan sizes are $500 \times 500\un{nm}^2$  with a resolution of 512 $\times$ 512 pixels$^2$.

Determining the packing fraction, $\phi = {\mdry}/{(\rhos \VT)}$, sheds light on the bulk ordering of the solids. The mass of a dried sample ($\mdry$), its total volume ($\VT$) and  the silica density ($\rhos$) are estimated as follows. To obtain $\mdry$, the sample are heated at $200\,^\circ\mathrm{C}$ for more than 3 hours to  remove  remaining water \cite{Ile79}. To obtain $\VT$, imbibition and Archimedes' principle are used: After immersion, the  pores soak up water and the sample's total volume is obtained via hydrostatic weighing\cite{Bar13}. This process was repeated using ethanol rather than water; the two values of $\VT$ agree within less than $2\%$. To obtain an accurate estimation of $\rhos$, the dilution method  is invoked \cite{FinMorBot85, Ada04}: Several diluted suspensions of Ludox, {of densities $\rho_{\rm diluted}$}, are prepared at prescribed mass fractions $(\phi^m_{\rm diluted})$. The initial mass fraction $\phi^m_0$ is estimated by weighing dry residues at $200\,^\circ\mathrm{C}$ and $\rhos$ is inferred by linear regression using the relationship $1-(\rho_{\rm water}/\rho_{\rm diluted}) = \phi^m_{\rm diluted}(1-(\rho_{\rm water}/\rhos))$. We found $\rhos = 2.26  \un{g/cm}^3$. Chemically adsorbed water on the nano-particles leads to a possible overestimation of $\mdry$ and underestimation of  $\rhos$, and thus, an overestimation of the packing fraction $\phi$. Silica dried at $200\,^\circ\mathrm{C}$ can retain up to 5 silanol groups per nm$^2$ \cite{Zhu93}. This corresponds to as much as $1.5\%$ residual water in mass, leading to atmost $3.6\%$ overestimation of $\phi$.

{\it Results--} Figures \ref{fig:rough}A and B show snapshots of the cracking pattern observed in the Petri dish  after drying (i.e. when the mass loss is negligible) for the two extremes. Drying quickly ($\RH=10\%$) leads to a large quantity of small fragments ($\sim 50$  visible fragments in the center of the Petri dish, within a disk of radius $\sim 1.75 \un{cm}$). On the contrary, the slowest drying rate ($\RH=95\%$) leads to a small quantity of large fragments (5 fragments visible over the same area). 
Generally, the typical fragment size decreases as $\edot_0$ increases (or equivalently as $\RH$ decreases).

The drying rate also alters the nanoscale arrangement of the colloids at the layer surface as observed via the AFM. An almost 2D crystalline structure forms for the fastest rate ($\RH=10 \%$; Fig. \ref{fig:rough}A'), and an amorphous one for the lowest rate ($\RH=95 \%$; Fig. \ref{fig:rough}B').

\begin{figure}[h!]
\includegraphics[width=\columnwidth]{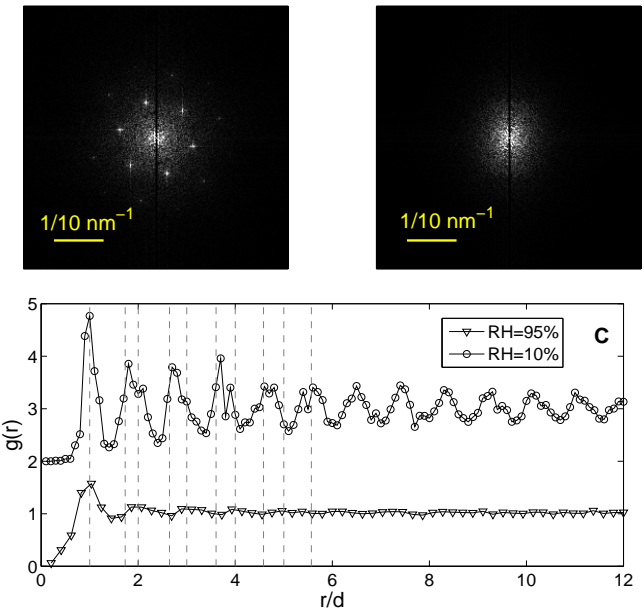}	
\caption{Fourier spectrum for $\RH=10\%$ 
(A) and $\RH=95\%$ (B), 
obtained by Fast Fourier transforms 
of the two topographical AFM images presented in Fig. \ref{fig:rough}. Note the sixfold spot symmetry for $\RH=10\%$, which is the signature of an hexagonal particle arrangement. The spots vanish for $\RH=95\%$, and the Fourier pattern become an isotropic disk, which is the signature of amorphous-like arrangement.  
Panel C shows the pair correlation function $g(r)$ for these two images. The radius has been normalized by the colloid diameter $d \simeq 15\un{nm}$ (as measured from  Fig. \ref{fig:rough}). The vertical dashes lines correspond to the peaks of a perfect hexagonal lattice: $r_1/d=1$, $r_1/d=\sqrt{3}$, $r_1/d=2$, $r_1/d=\sqrt{7}$, $r_1/d=3$, $r_1/d=\sqrt{13}$, $r_1/d=4, r_1/d=\sqrt{21}$, $r_1/d=5$, $r_1/d=\sqrt{31}$. For sake of clarity, the $g(r)$ obtained for $\RH=10\%$ is shifted upwards by two units.}
\label{fig:FFT}
\end{figure}
\nocite{DiGDavMit12, GauLazPau07}

Two-dimensional Fourier spectrum  conducted on the AFM topographical images shown in Figs. \ref{fig:FFT}A ($\RH=10\%$) and B ($\RH=95\%$)  access the differences in particle arrangements on the drying surface.  For the fastest evaporation rate, the pattern exhibits a sixfold symmetry characteristic of  a hexagonal  2D lattice. On the contrary, no discrete wavenumber can be identified for $\RH=95\%$ and only a central circular halo occurs. This is characteristic of an amorphous structure.

The pair correlation function, $g(r)$, represents  the probability of finding the center of a particle at a distance $r$ away from a given reference particle. Moreover, it aids in characterizing the long range order of the images. As in standard practice, $g(r)$ is normalized by that of an ideal gas (i.e. non-correlated particle positions) \cite{Hansen05}. Fig. \ref{fig:FFT}C depicts the two extreme cases: $\RH=10\%$ (top curve with circular points) and $\RH=95\%$ (bottom curve with triangular points). For $\RH=10\%$, $g(r)$ presents a well defined sequence of peaks extending more than $12 \,d$ where $d \simeq 15$ nm is the particle diameter acquired from Fig. \ref{fig:rough}A' and \ref{fig:rough}B' (note that this value is slightly higher than the spec one). The first several $g(r)$ peaks coincide well with theoretical values for a perfect hexagonal lattice (indicated by dash vertical lines),
even if they widen as $r$ increases. This is the signature of a quasi-long range translational order. Conversely, the $\RH=95\%$ peaks become barely visible for $r \geq 2d$ indicating a loss of translational order, as expected for amorphous structure.

\begin{figure}[h!]	
\includegraphics[width=\columnwidth]{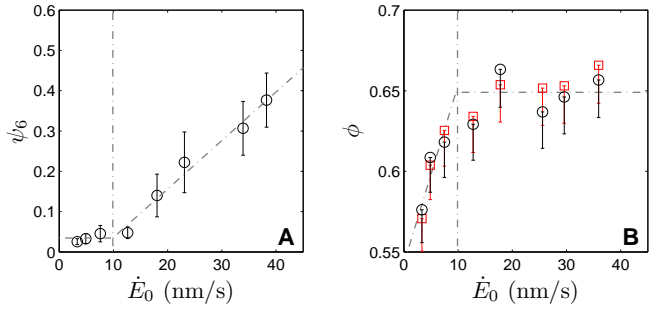}		
\caption{{A: Bond orientational order parameter ($\psi_6$) of the particle arrangement as a function of the evaporation rate $\edot_0$. Error bars indicate the standard deviation over measurements performed on 10 different AFM images. The dashed horizontal and inclined lines correspond to linear fits over the two rate regimes and the vertical line in between indicates the crossover value $\edot_c\simeq 10\un{nm/s}$. B: {Overall packing fraction $\phi$ as a function of $\edot_0$ measured by immersion of the samples in water (black circles) and ethanol (red squares). The error bars take into account 
the overestimation of $\phi$ due to residual water.}}} 
	\label{fig:single}
\end{figure}

The orientational order evolution can be investigated via the bond angle order parameter $\psi_n$ defined by 
\begin{equation}
\psi_n = \left| \frac{1}{M}\sum_{k=1}^{M} \frac{1}{N_k}\sum_{l=1}^{N_k} \exp{(i \, n \, \theta_{kl})}  \right|,
\end{equation}
where  
$n$ is the number of nearest neighbors,
$M$ is the number of particles in the AFM image, 
$N_k$ corresponds to the number of nearest neighbors of particle $k$, and 
$\theta_{kl}$  is the angle between a fixed direction and the  line linking the centers of particles $k$ and $l$ \cite{IngReiSha06}. 
For the six-fold symmetry visualized in Fig~\ref{fig:FFT}A, $n$ should be set to $6$.
For a perfect hexagonal arrangement $\psi_6 = 1$ while for an amorphous phase $\psi_6 =0$. 
The evolution of $\psi_{6}$ with $\edot_0$ (shown
in Fig. \ref{fig:single}A) exhibits two regimes with a crossover at $\edot_c \sim 10\un{nm/s}$.  For  $\edot_0 \leq \edot_c$, $\psi_6 \simeq 0.05\ll 1$. This reflects a disordered surface for low evaporation rates.  On the other hand when $\edot_0 > \edot_c$, $\psi_6$ increases linearly with $\edot_0$.
This is consistent with the observations of the translation order (fig. \ref{fig:FFT}C): Increasing the drying rate yields increasing hexagonal order at the surface. The fact that $\psi_6$ remains significantly lower than one reveals that even for the fastest drying rate the surfaces are not perfect 2D hexagonal lattices.

The above AFM analysis is restricted to {\it surface} characterizations of the particle arrangement.
In a complementary way, the particle volume fraction $\phi$ provides information on the {\it bulk} particle arrangement.
Fig. \ref{fig:single}B displays the evolution of $\phi$  as a function of $\edot_0$. Two regimes occur with a crossover $\edot_c \simeq 10\un{nm/s}$.  This crossover coincides with that observed in Fig. 3A which is related to the surface orientational order.  In the first regime $\edot_0 \leq \edot_c$, $\phi$ increases linearly with $\edot_0$. 
In the second regime $\edot_0 > \edot_c$, $\phi$ increase rapidly slows down and even saturates to $\phi \simeq 0.65 \pm 0.02 $, which is just above the random close packing value ($\phi_{RCP} = 0.64$ for monodisperse packing) but below the value of a compact structure ($\phi_{HCC/FCC} = 0.74$). This shows that the bulk is, to a large extent, disordered and  raises the question to what extend does the order extend into the bulk.

\begin{figure}[h!]
\includegraphics[width=\columnwidth]{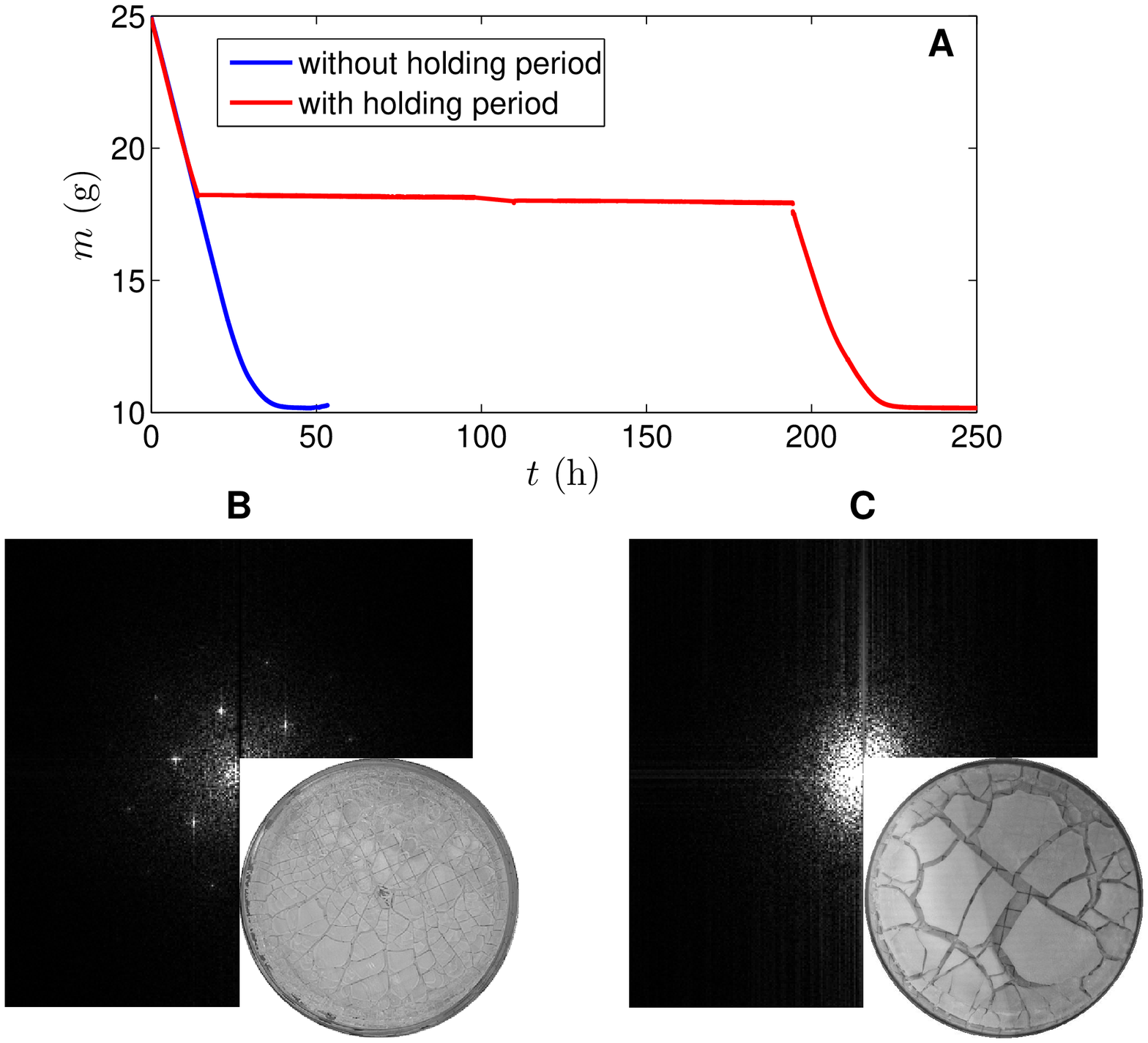}		
	\caption{Influence of the drying protocol on the dried layer's nanostructure and ability to crack. A: Time evolution of the Ludox mass $m$ within the Petri dish in a drying experiment performed at $\RH=10\%$ with and without a holding period. This holding period occurs when the solid fraction is $\phi=0.335$ and lasts $180\un{h}$. 
B: Fourier spectrum of a typical topographical AFM images (main panel) and crack pattern (inset) observed after evaporation without the holding period (this figure  is a composite of Fig. \ref{fig:FFT}A and Fig. \ref{fig:rough}A, and   facilitates the comparison with panel C). C: Fourier spectrum of a typical topographical AFM images (main panel) and crack pattern (inset) observed after evaporation with the holding period.}
	\label{fig:peclet}
\end{figure}

{\it Discussion--} 
The packing structure results from the way the particles bundle. Its dependency on the  drying rate is usually rationalized \cite{RouZim04, EkaKedMcD09} via the dimensionless P\'eclet number: 
$Pe \equiv h_{0} \edot_0/D_0$ 
where $D_{0}$ is the diffusion coefficient  for silica particles. Using Stokes-Einstein relation, one gets
$D_0 \simeq 2.04 \times 10^{-11}\,\mathrm{m}^2\mathrm{/s}$ at $T=25^\circ\mathrm{C}$. 
$Pe$ represents the ratio between the convection time for the particles toward the suspension surface and their Brownian diffusion time.
A high $Pe$ indicates directional packing implying the solid forms layer by layer.
A low $Pe$ means uniform and isotropic bulk compaction.

Checking whether or not $Pe$ controls the packing structure requires two additional experimental runs.  
The experiments invoke two parameter sets $\{h_0,\RH\}$ at a similar P\' eclet number well above $Pe = 1.5$ (i.e. $\edot_0 > \edot_c = 10\un{nm/s}$) which should result in a similar surface crystalline arrangement. 
The first experimental run uses $h_{0}=5$ mm 
and $\edot_{0}=36.1\un{nm/s}$  (RH=10 $\%$) and yields $Pe=4.4$. The second run uses $h_{0}=10$ mm
and $\edot_0=26\un{nm/s}$ (RH = 50 $\%$) and yields $Pe=6.4$ (slightly larger). The first run exhibits an ordered surface arrangement, while a disordered arrangement occurs for the second one.
This  demonstrates that $Pe$ is not the relevant parameter driving the particle packing.

Beyond the competition between convection and diffusion, the formation of aggregates within the suspension could be an important mechanism  driving the crystalline-to-amorphous arrangement in the solid. If this is the scenario, then slower drying favors the formation of aggregates and subsequently the formation of an amorphous solid. Testing this hypothesis calls for two new experimental runs. Both runs invoke $\RH=10 \%$ and $h_{0}=5$ mm (Fig.~\ref{fig:peclet}). The first experimental run uses, as previously, a continuous drying scheme and gives a crystalline surface arrangement. The second experimental run invokes an $180$ hours holding period (i.e. $\edot_{0} \sim 0$ g/h for $180$ hours) launched when $m = 18.2\un{g}$. During this holding period $\phi = 0.335$, the suspension is concentrated enough to favor agglomerate formation. After the holding period,  the sample resumes drying with $\RH=10\%$ until $\edot_0 \sim 0$. The solid layer exhibits an amorphous arrangement on the drying surface. Thus, it is conjectured that long drying protocols (including holding periods) favor the agglomerate formation which subsequently alters the nanostructure packing (amorphous instead of crystalline as imaged with the AFM). At the macroscale, the crack patterns of the second run are larger than the ones obtained, without a holding period, although they appear under the same external drying conditions. This implies that slowing down the drying enhances the macroscopic resistance to fracture by changing the way the particles bundle.

Conjecturing agglomerate formation sheds new light 
on  the $\psi_{6}$ {\em vs.} $\edot_{0}$ and $\phi$ {\em vs.} $\edot_{0}$ curves displayed  in fig. \ref{fig:single}.  At first glance, one expects a correlated increase of $\psi_{6}$ and $\phi$ with $\edot_{0}$  as they both increase with the packing order, yet this does not occur. Still, the curves can be understood assuming that 
$\edot_{c}$    is a critical value below which every particle of the suspension  is part of an aggregate  before solification. Subsequently:
\begin{itemize}
\item For $\edot_0 < \edot_{c}$,  the packing has to be amorphous and $\Psi_{6} \sim 0$.  
Increasing $\edot_{0}$ causes the mean size of the aggregates to decrease, hence the mean size of the pores decreases, {thus increasing the packing fraction} (i.e. $\phi$ increases).
\item For $\edot_0 > \edot_{c}$, the number of isolated particles increases with $\edot_{0}$, 
making it more likely to have crystalline zones, at least at the surface, where surface tension possibly drives the ordering \cite{BigCorJae06}.
 
As a result, $\psi_{6}$ increases with $\edot_{0}$. Additionally, isolated particles fill the space between the remaining aggregates.  This explains 
the slower increase, and even the saturation of $\phi$.
\end{itemize}

In conclusion, this paper demonstrates that the evaporation rate, during drying of a colloidal layer, modifies not only the drying kinetics but also the final dried material: Increasing the rate favors more ordered arrangements and denser packing. Explaining this scenario requires more than  the Peclet number and its simple balance between advection and diffusion. Studies herein suggest that aggregation as a result of the drying protocol plays a significant role in the formation of the solid. This may be of practical interest in the design of colloidal drying processes to obtain tunable and well-controlled 3D nanoparticle self-assemblies (e.g. crystals, modulated porosity solids...) over large dimensions, with innovative photonics and biotechnology applications \cite{DutHof04}. 
This rate dependent solid formation has a consequence on the fracture behavior: The resistance to fracture increases by decreasing the drying rate or by introducing a holding period at a wisely chosen time. Work in progress aims at quantifying how the material properties at the macroscale, notably its elastic modulus and toughness, emerge from the nanostructure and its history of formation. 

\acknowledgments
The co-authors would like to thank  F. Doumenc, B. Guerrier, L.-T. Lee,  P. Reis, J.-B. Salmon  for fruitful discussions,
A. Aubertin for the controlled drying chamber and for C. Wiertel-Gasquet for technical support in Labview. 
This research is supported by Triangle de la Physique (RTRA), Ile-de-France (C'Nano and ISC-PIF) and Investissements d'Avenir of LabEx PALM (ANR-10-LABX-0039-PALM) and LabEx LaSIPS (ANR-10-LABX-0040-LaSIPS).


\end{document}